\documentclass[a4paper,10pt]{article}
\pdfoutput=1 % if your are submitting a pdflatex (i.e. if you have
\usepackage{jheppub} % for details on the use of the package, please
\usepackage[T1]{fontenc} % if needed
\usepackage{lmodern}

\usepackage{tikz}
\usetikzlibrary{positioning,decorations.pathmorphing}

\let\oldFootnote\footnote
\newcommand\nextToken\relax

\renewcommand\footnote[1]{%
    \oldFootnote{#1}\futurelet\nextToken\isFootnote}

\newcommand\isFootnote{%
    \ifx\footnote\nextToken\textsuperscript{,}\fi}

\usepackage{enumerate}

\usetikzlibrary{decorations.markings}%-------------------------------------%

\def\id{{1 \kern-.28em {\rm l}}}

\def\K3{{\bf K3}}
\def\journal#1&#2(#3){\unskip, \sl #1\ \bf #2 \rm(19#3) }
\def\andjournal#1&#2(#3){\sl #1~\bf #2 \rm (19#3) }

\def\bar{\overline}

\def\ie{{\it i.e.}}
\def\eg{{\it e.g.}}

\def\tilde{\widetilde}

\def\frac#1#2{{#1\over#2}}

\def\inbar{\,\vrule height1.5ex width.4pt depth0pt}
\def\IC{\relax\hbox{$\inbar\kern-.3em{\rm C}$}}
\def\IR{\relax{\rm I\kern-.18em R}}
\def\IP{\relax{\rm I\kern-.18em P}}

\catcode`\@=11
\def\slash#1{\mathord{\mathpalette\c@ncel{#1}}}
\def\MM{{\cal M}}

\def\underrel#1\over#2{\mathrel{\mathop{\kern\z@#1}\limits_{#2}}}

\catcode`\@=12

\def\ie{{\it i.e.}}
\def\eg{{\it e.g.}}

\title{Strings in Irrelevant Deformations of $AdS_3/CFT_2$}

\author{}
\author{Soumangsu Chakraborty$^a$,}
\author{Amit Giveon$^b$}
\author{and David Kutasov$^c$}

\affiliation[a]{Department of Theoretical Physics,\\Tata Institute of Fundamental Research, Mumbai 400005, India}
\affiliation[b]{Racah Institute of Physics, The Hebrew University \\ Jerusalem 91904, Israel}
\affiliation[c]{EFI and Department of Physics, University of Chicago\\ 5640 S. Ellis Ave, Chicago, IL 60637, USA}

\abstract{We generalize our recent analysis \cite{Chakraborty:2020swe} of probe string dynamics to the case of general single-trace 
$T\bar T$, $J\bar T$ and $T\bar J$ deformations. We show that in regions in coupling space where the bulk geometry is smooth, the classical trajectories of such strings are smooth and approach the linear dilaton boundary at either the far past or the far future. These trajectories give rise to quantum scattering states with arbitrarily high energies. When the bulk geometry has closed timelike curves (CTC's), the trajectories are singular for energies above a critical value $E_c$. This singularity occurs in the region with CTC's, and the value of $E_c$ agrees with that read off from the dual boundary theory for all values of the couplings and charges.}

\begin{document}
\maketitle
\flushbottom

%\section{Introduction and Summary}

In this letter, we continue the study \cite{Giveon:2017nie,Giveon:2017myj,Asrat:2017tzd,Giribet:2017imm,Chakraborty:2018kpr,Chakraborty:2018vja,Apolo:2018qpq,Chakraborty:2018aji,Chakraborty:2019mdf,Apolo:2019yfj,Hashimoto:2019wct,Hashimoto:2019hqo,Apolo:2019zai,Asrat:2020uib,Chakraborty:2020xyz,Chakraborty:2020swe} of a class of solvable irrelevant deformations of $AdS_3/CFT_2$ duality.
These models touch on a number of important issues in quantum field theory and string theory.

On the field theory side, they provide examples of theories that do not approach at high energies a fixed point of the renormalization group (a CFT). In some cases, they give rise to a Hagedorn density of states. In others, the spectrum of energies is bounded from above, and the total number of states (on a circle) is finite. In both cases, the resulting physics is non-local, and it is interesting to study it in a controlled setting.

On the string theory side, these theories are described by backgrounds \cite{Forste:1994wp,Giveon:1999zm,Israel:2004vv,Detournay:2010rh,Chakraborty:2019mdf} that are not asymptotically AdS. When the high-energy spectrum is Hagedorn, the corresponding background is asymptotically flat spacetime with a linear dilaton. Thus, the resulting theory is a two-dimensional vacuum of Little String Theory (LST) \cite{Aharony:1998ub,Aharony:1999ks,Kutasov:2001uf}.  When the energy is bounded from above, the background looks sick, containing some combination of naked singularities and
regions with closed timelike curves (CTC's).

Conservatively, one would discard such backgrounds, but in this context they are obtained by a seemingly well behaved deformation
of a unitary theory (string theory on $AdS_3$). Moreover, in some cases, the original $AdS_3$ vacuum preserves an extended supersymmetry,
and the deformation in question preserves the full supersymmetry (but, of course, not the conformal symmetry, and in general also not
the Lorentz symmetry). Thus, it is interesting to understand whether the resulting string theory backgrounds (and corresponding boundary
theories) are consistent, despite appearances, or if they are not, identify the inconsistency.

Viewing these theories from the point of view of holographic duality leads to an interesting interplay between features of the boundary theory (such as the spectrum, thermodynamics, correlation functions of local operators and entanglement entropy), and those of the bulk one (such as the singularity structure of the bulk spacetime, and properties of black holes that dominate the thermodynamics). This is an important motivation for studying them.

The starting point of the construction of these models is string theory on $AdS_3$ supported by a Neveu-Schwartz B-field. While the construction is rather general, to have a concrete model in mind one can focus on the special case $AdS_3\times S^3\times T^4$, which is the near-horizon geometry of $k$ NS5-branes and $p$ fundamental strings. The number of fivebranes determines the level of the $SL(2,\mathbb{R})$ current algebra of the sigma model on $AdS_3$ and the $SU(2)$ current algebra of the sigma model on $S^3$. The number of strings determines the string coupling, $g_s^2\sim 1/p$. Thus, we take $p$ to be large, so that the corresponding string coupling is small.

 It is known that the string theory described above is dual to a two-dimensional CFT \cite{Maldacena:1997re}, but that CFT is not understood in detail. One thing that is known about it is that its spectrum of scaling dimensions contains a continuum above a finite gap. This continuum is due to long strings \cite{Seiberg:1999xz}, which can be thought of as some of the $p$ strings that make the $AdS_3$ background, carrying arbitrary radial momentum. Since these strings are free (to leading order in $1/p$) in a wide range of radial positions, their spectrum is described by a symmetric product CFT \cite{Argurio:2000tb,Giveon:2005mi}; see \cite{Chakraborty:2019mdf} for a recent discussion and further references. In a sector with $w$ long strings, the spectrum is that of the symmetric product,
\begin{eqnarray}
\left(\MM_{6k}^{(L)}\right)^w/S_w,\label{symmm}
 \end{eqnarray}
where $\MM_{6k}^{(L)}$ is the theory on a single long string described in \cite{Seiberg:1999xz}.

When the long strings approach the boundary, the coupling on them increases \cite{Seiberg:1999xz}, and eventually one loses perturbative control over them. In this regime, they are better described as BTZ black holes, and a priori one might expect that the symmetric product description \eqref{symmm} breaks down. One of the outcomes of the analysis of previous papers \cite{Giveon:2017nie,Chakraborty:2020swe,Apolo:2019zai}
is that the black holes appear to be well described by a symmetric product,
with the building block $\MM_{6k}^{(L)}$ replaced by one with more degrees of freedom, $\MM_{6k}$.

An important recent development in two-dimensional quantum field theory is the realization that two-dimensional CFT's have a universal irrelevant
deformation that allows one to flow up the renormalization group -- the $T\bar T$ deformation \cite{Smirnov:2016lqw,Cavaglia:2016oda}. In general, irrelevant deformations
of a CFT do not lead to a well defined RG flow to the UV, but in this case, one can define the theory on all scales, by using the fact that
it has a conserved stress tensor for all values of the coupling, and using it to flow up the RG.\footnote{To be precise,
this procedure is valid to all orders in the coupling; non-perturbatively, one may encounter new issues.}

String theory on $AdS_3$ has a similar, but distinct, deformation \cite{Giveon:2017nie}.
This deformation owes its existence to the presence in the dual CFT of a single-trace operator $D(x)$ with (left,right) scaling dimension
$(2,2)$. Single-trace means in this context that it is described in the worldsheet theory by a (non-normalizable) vertex operator integrated over the worldsheet. This operator has much in common with the operator $T\bar T$ in the boundary CFT,
however, the latter is a product of two integrated vertex operators \cite{Kutasov:1999xu}.

Adding to the Lagrangian of the boundary CFT the operator $D(x)$ corresponds in the worldsheet theory to a current-current deformation of the $SL(2,\mathbb{R})$ sigma model \cite{Giveon:2017nie}. This deformation is exactly marginal on the worldsheet, and leads to a solvable theory. This situation should be contrasted with adding to the Lagrangian of the boundary CFT a generic irrelevant operator, which leads in the bulk worldsheet description to the same sort of ambiguities familiar from irrelevant deformations in field theory.

The resulting bulk background is not asymptotically $AdS_3$ in the UV, which is consistent with the fact that the corresponding perturbation of the boundary theory is irrelevant. For one sign of the coupling of $D(x)$, which we call positive, one finds a smooth background that interpolates between $AdS_3$ in the IR and a linear dilaton spacetime in the UV. For the other, there is a singularity at a finite value of the radial coordinate, beyond which the signature of $1+1$ dimensional spacetime flips.

The above description of the perturbed theory is in terms of the bulk variables. It is natural to ask what is its description in terms of the boundary ones. Since we do not understand the structure of the boundary CFT well, we do not have a full answer to this question. However, in the long-string sector of the theory \eqref{symmm}, one can show that the deformation corresponding to $D(x)$ is a $T\bar T$ deformation of the block $\MM_{6k}^{(L)}$ \eqref{symmm}. Indeed, the spectrum of long strings in the deformed theory is given in terms of the undeformed spectrum by the same expression as that found in \cite{Smirnov:2016lqw,Cavaglia:2016oda}, applied to a symmetric product of $T\bar T$ deformed CFT's; see \eg\ \cite{Hashimoto:2019hqo} for a discussion of such symmetric products.

In particular, one finds that the two signs of the coupling give rise to different spectra. For positive coupling, one finds a spectrum that depends smoothly on the undeformed energy for arbitrary size of the latter, while for negative coupling, if the undeformed energy exceeds a certain critical value, the deformed one is complex. To investigate the fate of such states in the bulk theory, we recently \cite{Chakraborty:2020swe} studied the trajectory of a classical fundamental string moving radially in the deformed bulk geometry. We found that for positive coupling the trajectory of the string is smooth for all values of its radial momentum. Quantization of this trajectory leads to the states described by the equations of \cite{Smirnov:2016lqw,Cavaglia:2016oda}.

On the other hand, for negative coupling we found that when the radial momentum exceeds the value read off from the $T\bar T$ analysis, the trajectory of the string encounters a singularity. Interestingly, that singularity is not the curvature singularity of the geometry -- the string passes smoothly through that singularity. Rather, after it passes through the singularity of the background, the string encounters a singularity in its trajectory, whose location depends on the radial momentum, or equivalently the undeformed energy of the string. Thus, it is natural to expect that such trajectories do not give upon quantization physical states in the theory, in agreement with the expectations from $T\bar T$ deformed CFT.

The original construction of $T\bar T$ deformed CFT \cite{Smirnov:2016lqw,Cavaglia:2016oda} and its single-trace string theory realization \cite{Giveon:2017nie} have been generalized to a larger class of theories with multiple couplings. On the field theory side, this involves studying theories with holomorphic conserved currents $J(z)$, $\bar J(\bar z)$, and adding to the Lagrangian terms proportional to $J\bar T$, $T\bar J$, etc. The spectrum of these theories was found in \cite{Chakraborty:2018vja,Guica:2017lia,Chakraborty:2019mdf,LeFloch:2019rut}.

On the string theory side, the construction of \cite{Giveon:2017nie} was generalized to this larger set of theories in \cite{Chakraborty:2018vja,Apolo:2018qpq,Chakraborty:2019mdf}, where it was shown that the corresponding string backgrounds are obtained from $AdS_3\times \MM$, with $\MM$ a compact worldsheet CFT which contains some conserved currents, by a current-current deformation, generalizing that of \cite{Giveon:2017nie}.

Much of the discussion of the $T\bar T$ case can be extended to this larger set of theories. In particular, one can show that in the
long-string sector, the general current-current deformation of the worldsheet theory must correspond to a combined $T\bar T$, $J\bar T$, $T\bar J$ deformation of the block of the symmetric product \eqref{symmm}. Thus, the agreement of the spectrum of long strings in the current-current deformed backgrounds $AdS_3\times \MM$ and the spectrum of the corresponding symmetric product \eqref{symmm} with the corresponding deformation of the block $\MM_{6k}^{(L)}$ must generalize to this case. This was indeed shown to be the case in \cite{Chakraborty:2019mdf}.

It is therefore interesting to extend the discussion of \cite{Giveon:2017nie,Chakraborty:2020swe} to the more general space of theories. One motivation for doing that is to further study the interplay between features of the spectrum of the boundary theories obtained by the above irrelevant deformations, and features of the dual bulk geometries. As discussed \eg\ in \cite{Chakraborty:2019mdf}, the qualitative features of the bulk background depend on the region in coupling space we are in. In some regions in coupling space the background has naked singularities, while in others there are no singularities, but there are closed timelike curves in some region in the bulk spacetime. In both of these cases, the spectrum of states is truncated in the UV, and it is interesting to understand how the classical trajectories of long strings depend on the various features in the geometry.

Studying the above space of deformations is also useful for understanding better the original $AdS_3/CFT_2$ duality. This is a general theme in quantum field theory -- a way to understand a theory better is to study its space of deformations. In our case, the analysis of this and previous papers gives strong additional support to the statement that long strings on $AdS_3$ are described by a symmetric product CFT \eqref{symmm}.

In the rest of this letter, we study the classical trajectories of probe strings in the full class of geometries holographically dual to a linear combination of single-trace $T\bar T$, $J\bar T$ and $T\bar J$ deformed theories. We find that the picture of \cite{Chakraborty:2020swe}, obtained when only the $T\bar T$ coupling is turned on, generalizes to the full space of theories. In particular, the trajectories of strings have singularities iff the bulk geometry in which the strings propagate has a region with CTC's, and these singularities occur in that region. Moreover, the trajectories are singular iff the energy of the probe string exceeds the threshold above which the deformed energy in the corresponding deformed CFT is complex.

We also briefly comment on our expectations regarding properties of black holes in these geometries, and the corresponding states in the dual theory.

\medskip

The string background corresponding to a general linear combination of single-trace $T\bar{T}$, $J\bar{T}$ and $T\bar{J}$ deformations is described \cite{Chakraborty:2019mdf} by the sigma-model action
\begin{eqnarray}
S=\frac{1}{2\pi l_s^2}\int d^2z\left(\partial\phi\bar{\partial}\phi+h\partial\bar{\gamma}\bar{\partial}\gamma+2\epsilon_+ h\partial y\bar{\partial}\gamma+2\epsilon_-h\partial \bar{\gamma}\bar{\partial}y+\frac{h}{f}\partial y\bar{\partial}y\right),\label{defsig}
\end{eqnarray}
with
\begin{eqnarray}\label{fh}
\begin{split}
&f^{-1}=\lambda+e^{-\frac{2\phi}{\sqrt{k}l_s}},\\
&h^{-1}=\lambda-4\epsilon_+\epsilon_-+e^{-\frac{2\phi}{\sqrt{k}l_s}}.
\end{split}
\end{eqnarray}
The couplings $(\lambda,\epsilon_\pm)$ parametrize the space of $T\bar{T}$, $J\bar{T}$ and $T\bar{J}$ deformations. The coordinates $(\gamma,\bar{\gamma})$ label the $1+1$ dimensional boundary  $\mathbb{R}\times S^1$. They are related to the usual coordinates $(t,x)$ by
\begin{eqnarray}\label{ggb}
\gamma=x+t, \ \ \ \  \bar{\gamma}=x-t, \ \ \ \ x\sim x+2\pi R .
\end{eqnarray}
In addition to \eqref{defsig}, which describes the metric and $B$-field, there is also a non-trivial dilaton field,
\begin{eqnarray}\label{dil}
e^{2\Phi}=g^2 e^{-\frac{2\phi}{\sqrt{k}l_s}}h,
\end{eqnarray}
where $g$ is the string coupling in the infrared limit $\phi\to-\infty$, where the background \eqref{defsig}--\eqref{dil} approaches $AdS_3\times S^1$.

The $S^1$ labeled by $y$ is necessary for the existence of the worldsheet and spacetime $U(1)_L$, $U(1)_R$ currents that play a role in the construction. For example, in $AdS_3\times S^3\times T^4$ one can take $y$ to be one of the directions on $T^4$, or an $S^1\subset S^3$. The former has the advantage that the corresponding $J\bar T$ and $T\bar J$ deformations preserves supersymmetry.

As discussed in earlier papers \cite{Chakraborty:2019mdf,Chakraborty:2020xyz}, the background \eqref{defsig}--\eqref{dil}, describes a smooth spacetime with no pathologies in some region in coupling space, while in other regions it contains naked singularities and closed timelike curves. It appears that the existence of an upper bound on the energies of physical states is associated with the latter. Therefore, it will be useful for the subsequent discussion to analyze the region in coupling space and in spacetime where CTC's appear.

To do that, we focus on the compact coordinates $(x,y)$ and set the non-compact ones $(t,\phi)$ to constant values. The metric \eqref{defsig} takes in this case the form
\begin{eqnarray}\label{2dmetric}
ds^2=hdx^2+2h(\epsilon_++\epsilon_-)dxdy+\frac{h}{f}dy^2.
\end{eqnarray}
We want to find the conditions on the metric \eqref{2dmetric} such that all closed curves in the space labeled by $(x,y)$ are spacelike. Setting $y$ to a constant, the curve going around the $x$ circle is spacelike iff $h>0$, \ie
\begin{eqnarray}\label{condone}
\lambda-4\epsilon_+\epsilon_-+e^{-\frac{2\phi}{\sqrt{k}l_s}}>0.
\end{eqnarray}
If \eqref{condone} is satisfied, we can divide the metric \eqref{2dmetric} by the positive quantity $h$, and in order for all the remaining closed curves on the two-torus labeled by $(x,y)$ to be spacelike, we need the metric
\begin{eqnarray}\label{M}
M=\begin{pmatrix}
1 & \epsilon_++\epsilon_-\\
\epsilon_++\epsilon_- & \lambda+e^{-\frac{2\phi}{\sqrt{k}l_s}}
\end{pmatrix}
\end{eqnarray}
to be positive definite.

The eigenvalues of $M$ are given by
\begin{eqnarray}\label{eigenv}
m_\pm=\frac{1}{2}e^{-\frac{2\phi}{\sqrt{k}l_s}}\left(1+(1+\lambda)e^{\frac{2\phi}{\sqrt{k}l_s}}\pm\sqrt{4e^{\frac{4\phi}{\sqrt{k}l_s}}(\epsilon_++\epsilon_-)^2+\left(e^{\frac{2\phi}{\sqrt{k}l_s}}(\lambda-1)+1\right)^2}\right).
\end{eqnarray}
$m_+$ is always positive, but $m_-$ can be negative. To analyze its positivity properties, it is convenient to define the combination of couplings
\begin{equation}\label{A}
\Psi=\lambda-(\epsilon_++\epsilon_-)^2,
\end{equation}
following \cite{Chakraborty:2019mdf}. For $\Psi>0$, $m_-$ is always positive, in agreement with the statement that in this case the spacetime \eqref{defsig} is regular \cite{Chakraborty:2019mdf}. For $\Psi<0$, the requirement that this eigenvalue is positive reduces to
\begin{eqnarray}\label{cond}
e^{-\frac{2\phi}{\sqrt{k}l_s}}>|\Psi|.
\end{eqnarray}
Thus, for $\Psi<0$, CTC's appear beyond a certain critical value of $\phi$, given by \eqref{cond} with inequality replaced by equality. Note that the condition \eqref{cond} is more stringent than \eqref{condone} for all $(\lambda,\epsilon_\pm)$; thus, we can ignore the latter. Also, we only analyzed the condition that CTC's do not appear. In principle, it could be that in some cases other pathologies, such as naked singularities, appear in a region where \eqref{cond} is satisfied, but a study of subspaces of the full coupling space suggests that this does not happen.
Below we will make contact with \eqref{cond} by studying singularities of trajectories of probe strings in the background \eqref{defsig}--\eqref{dil}.

 We would like to generalize the discussion of probe strings in single-trace $T\bar T$ deformed CFT in \cite{Chakraborty:2020swe} to the class of theories with general couplings $\lambda$, $\epsilon_\pm$. As discussed above, perturbative long string states in the original CFT are described by the symmetric product \eqref{symmm}, and the above couplings can be shown to parametrize the space of deformations of the block CFT $\MM_{6k}^{(L)}$ by the operators $T\bar T$, $J\bar T$, $T\bar J$. Therefore, the energies of long strings in the background \eqref{defsig}--\eqref{dil} are described by the formula derived for these deformations in \cite{Chakraborty:2019mdf}. In particular, for $\Psi<0$, it is known that the energies of states become complex above a certain critical value. For the $T\bar T$ case, it was shown in \cite{Chakraborty:2020swe} that states with energies above this critical value correspond to singular classical trajectories, and thus it is natural to expect that they do not give rise to stable physical states in the deformed theory.

We will see that this is the case in general as well. Following \cite{Chakraborty:2020swe}, we will do this in two steps. First, we will perform a semiclassical quantization of strings in the background \eqref{defsig}. The main purpose of this is to (re)derive the energy formula in a form that will be useful for discussing the classical trajectories and, in particular, to calculate the upper bound on the energy in our conventions.

We will then analyze the classical trajectories that give rise to the semiclassical states mentioned above, and show that the upper bound read off from the spectrum is the same as the upper bound obtained from the study of the trajectories, as in \cite{Chakraborty:2020swe}. We will also show that the singularities of the trajectories are restricted to the region where \eqref{cond} is violated, \ie\ the region with CTC's.

%\noindent{\bf Semiclassical analysis of the spectrum}

As in \cite{Chakraborty:2018vja,Chakraborty:2020swe}, we parametrize the worldsheet cylinder by $-\infty<\tau<\infty$ and $\sigma\simeq\sigma +2\pi$,
\begin{eqnarray}
z=\frac{1}{\sqrt{2}}(\tau+\sigma), \ \ \ \ \  \bar{z}=\frac{1}{\sqrt{2}}(\tau-\sigma),\label{zzbst}
\end{eqnarray}
\begin{eqnarray}
\partial=\frac{1}{\sqrt{2}}(\partial_\tau+\partial_\sigma), \ \ \ \ \  \bar{\partial}=\frac{1}{\sqrt{2}}(\partial_\tau-\partial_\sigma).\label{der}
\end{eqnarray}
The conjugate momentum densities for the fields $(\phi,\gamma,\bar{\gamma},y)$ are given  by
\begin{eqnarray}
&&\Pi_\phi=T\dot{\phi},\label{piphig}\\
&&\Pi_\gamma=\frac{T}{\sqrt{2}}h\partial \bar{\gamma}+T\sqrt{2}\epsilon_+ h\partial y=\frac{T}{2}h(\dot{ \bar{\gamma}}+\bar{\gamma}')+T\epsilon_+ h(\dot{y}+y'),\label{pigammag}\\
&& \Pi_{\bar{\gamma}}=\frac{T}{\sqrt{2}}h\bar{\partial} \gamma+T\sqrt{2}\epsilon_- h\bar{\partial} y=\frac{T}{2}h(\dot{ \gamma}-\gamma')+T\epsilon_-h(\dot{y}-y'),\label{pigammabg}\\
&&\Pi_y =\sqrt{2}T\epsilon_+ h\bar{\partial}{\gamma}+\sqrt{2}T\epsilon_- h\partial\bar{\gamma}+\frac{Th}{\sqrt{2}f}(\partial y + \bar{\partial}y) \nonumber \\
&& \ \ \ \ \ =T\epsilon_+ h(\dot{ \gamma}-\gamma')+T\epsilon_- h(\dot{ \bar{\gamma}}+\bar{\gamma}') +T\frac{h}{f}\dot{y}, \label{piyg}
\end{eqnarray}
where, as usual, dot and prime denote derivatives with respect to the worldsheet coordinates $\tau$ and $\sigma$,
 respectively, and $T=1/2\pi l_s^2$ is the string tension.

 The worldsheet Hamiltonian and momentum are given by
 \begin{eqnarray}
 \begin{split}\label{hp}
 H&=&\frac{1}{2\pi}\int d^2\sigma\mathcal{H},\\
 p&=&\frac{1}{2\pi}\int d^2\sigma\mathcal{P},
 \end{split}
 \end{eqnarray}
 where the Hamiltonian and momentum densities are given by
 \begin{eqnarray}
 &&\mathcal{H}=\Pi_\phi\dot{\phi}+\Pi_\gamma\dot{\gamma}+\Pi_{\bar{\gamma}}\dot{\bar{\gamma}}+\Pi_y\dot{y}-\mathcal{L},\label{H}\\
 && \mathcal{P}=\Pi_\phi\phi'+\Pi_\gamma\gamma'+\Pi_{\bar{\gamma}}\bar{\gamma}'+\Pi_yy'.\label{P}
 \end{eqnarray}
 Let us consider a closed string at some radial position $\phi$, which carries momentum $P_\phi$ in the $\phi$ direction,
and let $E$ be the energy of the string conjugate to $t$ and $P$ the momentum conjugate to $x$.
 Since $x$ is compact, $P$ is quantized,
 \begin{eqnarray}
 P={n\over R}~,
 \end{eqnarray}
 where $n$ is an integer. The quantized momentum and winding quantum numbers along the $y$ direction will be denoted by $n_y$ and $m_y$,
 respectively. These spacetime quantum numbers are given by
\begin{eqnarray}
\int_0^{2\pi}d\sigma\Pi_{\gamma}&=&-\frac{E_L}{R},\label{ipigamma}\\
\int_0^{2\pi}d\sigma\Pi_{\bar{\gamma}}&=&\frac{E_R}{R},\label{ipigammab}\\
\int_0^{2\pi}d\sigma\Pi_{\phi}&=&P_{\phi},\label{ipiphi}\\
\int_0^{2\pi}d\sigma\Pi_{y}&=&\frac{n_y}{R_y}.\label{ipiy}
\end{eqnarray}
and
\begin{eqnarray}
&&\int_0^{2\pi}d\sigma \gamma'=\int_0^{2\pi}d\sigma\bar{\gamma}'=2\pi w R,\label{pqn1} \\
&&\int_0^{2\pi}d\sigma y'=2\pi m_y R_y,\label{pqn2}\\
&&\int_0^{2\pi}d\sigma\phi'=0,\label{pqn3}
\end{eqnarray}
where
\begin{eqnarray}
\begin{split}\label{EP}
E_L&=&\frac{R}{2}(E+P),\\
E_R&=&\frac{R}{2}(E-P),
\end{split}
\end{eqnarray}
and $w$ is the winding of the string in the $x$ direction.

The contribution of the string zero modes to the worldsheet Hamiltonian $H$, denoted by $ \Delta+\bar{\Delta}$, and to the momentum $p$, denoted by $ \bar{\Delta}-\Delta$,  are
\begin{eqnarray}
\begin{split}\label{ddb}
\Delta+\bar{\Delta}&=-w(E_L+E_R)-\frac{2l_s^2}{R^2f}E_LE_R+\frac{1}{2}(q_L^2+q_R^2)+\frac{2l_s^2}{R^2}(\epsilon_-E_L+\epsilon_+E_R)^2\\
&+\frac{2\sqrt{2}l_s}{R}(\epsilon_-q_RE_L+\epsilon_+q_LE_R) +\frac{l_s^2 P_\phi^2}{2},\\
\bar{\Delta}-\Delta&=\frac{1}{2}\left(q_R^2-q_L^2\right)+nw,
\end{split}
\end{eqnarray}
 where the charges $q_{L,R}$ are given by
 \begin{eqnarray}
 q_{L,R}=\frac{1}{\sqrt{2}}\left(\frac{n_yl_s}{R_y}\pm\frac{m_yR_y}{l_s} \right)~.
 \end{eqnarray}
For studying the relation with irrelevant deformations of a CFT, we set $w=1$ (states with winding $w>1$ correspond to twisted sectors of the symmetric product of deformed CFT's \cite{Argurio:2000tb,Chakraborty:2019mdf,Hashimoto:2019hqo}).

As an example, we restrict to states with zero momentum in the $x$ direction, \ie\  $n=0$, and no right-moving excitations.  The Virasoro constraints are $\Delta+N=\bar\Delta=0$, where $N=\frac{1}{2}(q_R^2-q_L^2)$ is a non-negative integer,
equal to the left-moving excitation level on the string.

As in \cite{Chakraborty:2018vja,Chakraborty:2020swe}, the Virasoro constraints lead to the dispersion relation
\begin{eqnarray}\label{ubeg}
 E=E_{\rm c}+{\rm sign(\Psi)}\sqrt{E_{\rm c}^2+\frac{2q_R^2}{l_s^2\Psi}+\frac{P_\phi^2}{\Psi}},\label{mshellg}
 \end{eqnarray}
 where
 \begin{eqnarray}
 E_{\rm c}=\frac{\sqrt{2}\epsilon_+q_L}{l_s\Psi}+\frac{\sqrt{2}\epsilon_-q_R}{l_s\Psi}-\frac{R}{l_s^2\Psi} .\label{ecg}
 \end{eqnarray}
 For positive $\Psi$, the energy \eqref{ubeg} is not bounded from above. It is bounded from below by zero, in agreement with the fact that the states we constructed are scattering states, whose energy must be above the vacuum. In some cases the theory is supersymmetric, and $E>0$ is required by unitarity.

 For negative $\Psi$, one can think of $E_c$ \eqref{ecg} as the maximal energy in the theory. Superficially, it looks like by changing the couplings and charges it can be made negative, but it is straightforward to see that before one gets to that point, all the states in the particular charge sector develop complex energies, and (as we will see next) do not exist as physical states in the theory.

In \cite{Chakraborty:2020swe}, we showed that for single-trace $T\bar T$ deformed CFT (\ie\ for $\epsilon_\pm=0)$, states for which the energy \eqref{ubeg} is complex correspond to singular classical trajectories and, moreover, the singularity of these trajectories is located in the region where the corresponding geometry \eqref{defsig}--\eqref{dil} has closed timelike curves.

Here, we generalize this picture to the full coupling space labeled by $(\lambda,\epsilon_\pm)$. To make contact with the discussion above, we will take the string to wrap the $x$ circle once and the $y$ circle $m_y$ times, and solve for its radial trajectory $\phi(t)$. As there, we will take the momentum of the string in the $x$ direction, $P$, to be zero.

The equations of motion obtained from varying the action \eqref{defsig} are given by
\begin{eqnarray}
&&\bar{\partial}\partial\phi+h'\left(\partial\bar{\gamma}\bar{\partial}\gamma+2\epsilon_+ \partial y\bar{\partial}\gamma+2\epsilon_-\partial \bar{\gamma}\bar{\partial}y\right)+\left(\frac{h}{f}\right)'\partial y\bar{\partial}y=0,\label{a}\\
&& \partial\left(h\bar{\partial}\gamma+2\epsilon_- h\bar{\partial} y\right)=0,\label{b}\\
&& \bar{\partial}\left(h\partial \bar{\gamma}+2\epsilon_+ h\partial y \right)=0,\label{c}\\
&&\partial \left( \epsilon_+h \bar{\partial}\gamma \right)+\bar{\partial} \left( \epsilon_-h \partial\bar{\gamma} \right)+\frac{h}{f}\bar{\partial}\partial y=0\label{d}.
\end{eqnarray}
Fixing the residual reparametrization symmetry of the action \eqref{defsig}, eqs. \eqref{b} and \eqref{c} take the form
\begin{eqnarray}
\begin{split}
&h\bar{\partial}\gamma+2\epsilon_- h\bar{\partial} y=c, \label{eqag}\\
& h\partial \bar{\gamma}+2\epsilon_+ h\partial y=-c,
\end{split}
\end{eqnarray}
where $c$ is a constant given by
\begin{eqnarray}
c=\frac{E}{2\sqrt{2}\pi T}.\label{con}
\end{eqnarray}

The holomorphic and anti-holomorphic components of the stress tensor are given by
\begin{eqnarray}
\begin{split}\label{ttbg}
&T_{zz}=\frac{1}{2l_s^4}\left((\partial \phi)^2+h\partial \gamma\partial \bar{\gamma}+2\epsilon_+ h\partial y\partial \gamma +2\epsilon_-h\partial y \partial \bar{\gamma}+\frac{h}{f}(\partial y)^2\right),\\
&T_{\bar{z}\bar{z}}=\frac{1}{2l_s^4}\left((\bar{\partial} \phi)^2+h\bar{\partial} \gamma\bar{\partial} \bar{\gamma}+2\epsilon_+ h\bar{\partial} y\bar{\partial} \gamma +2\epsilon_-h\bar{\partial} y \bar{\partial} \bar{\gamma}+\frac{h}{f}(\bar{\partial} y)^2\right).
\end{split}
\end{eqnarray}
Using the equations of motion \eqref{a} -- \eqref{d}, one can check the conservation of the stress tensor, $\bar{\partial} T_{zz}=\partial T_{\bar{z}\bar{z}}=0$.

Using \eqref{piyg}, \eqref{ipiy} and \eqref{pqn2}, one can write
\begin{eqnarray}\label{ydypg}
\dot{y}=Y-\sqrt{2}c(\epsilon_+-\epsilon_-), \ \ \ \ y'=\tilde{c},
\end{eqnarray}
 where $Y$ and $\tilde{c}$ are given by
  \begin{eqnarray}
Y=\frac{n_y}{2\pi T R_y}, \ \ \ \  \tilde{c}=m_yR_y.\label{Xct}
 \end{eqnarray}
From \eqref{pqn3} and the fact that we are considering a string with no transverse oscillator excitations in the radial direction, it follows that  $\phi=\phi(\tau)$. The Virasoro constraints imply
\begin{eqnarray}
T_{zz}-T_{\bar z\bar z}&=&\frac{1}{2l_s^2}\left(q_L^2-q_R^2\right),\label{vira}\\
T_{\bar z\bar z}&=&0\label{virb}.
\end{eqnarray}
From \eqref{vira}, one obtains
\begin{eqnarray}
\begin{split}\label{xdot}
&\dot{x}=\tilde{c}\left(\frac{\sqrt{2}Y}{c}+\epsilon_--\epsilon_+\right)- \left(Y+\sqrt{2}c(\epsilon_--\epsilon_+)\right)(\epsilon_++\epsilon_-) -\frac{l_s^2}{\sqrt{2}c}\left(q_L^2-q_R^2\right).
\end{split}
\end{eqnarray}
This leads to the following ansatz:
\begin{eqnarray}\label{xt}
&&x=R\sigma+\tilde{c}\left(\frac{\sqrt{2}Y}{c}+\epsilon_--\epsilon_+\right)\tau - \left(Y-\sqrt{2}c(\epsilon_+-\epsilon_-)\right)(\epsilon_++\epsilon_-)\tau-\frac{l_s^2}{\sqrt{2}c}\left(q_L^2-q_R^2\right)\tau,\nonumber \\
&&t=t(\tau).
\end{eqnarray}
The other Virasoro constraint, \eqref{virb}, gives
\begin{eqnarray}\label{phid}
\dot{\phi}^2-\sqrt{2}c(\dot{t}-\dot{x}+R+2\tilde{c}\epsilon_--2Y\epsilon_-)-2c(\epsilon_+^2-\epsilon_-^2)+(Y-\tilde{c})^2=0.
\end{eqnarray}
From  \eqref{eqag}, \eqref{xdot} and \eqref{phid}, one obtains
\begin{eqnarray}
\resizebox{.92\hsize}{!}{$\dot{t} =\sqrt{2}c\left[e^{-\frac{2\phi}{\sqrt{k}l_s}}+\frac{1}{\sqrt{2}c}\Big(\sqrt{2}c\Psi+R+ \tilde{c}(\epsilon_++\epsilon_-)+Y(\epsilon_+-\epsilon_-)-\frac{\sqrt{2}\tilde{c}Y}{c}+\frac{l_s^2}{\sqrt{2}c}\left(q_L^2-q_R^2\right)\Big)\right],$}
\label{solg1}\\
\resizebox{.92\hsize}{!}{$\dot{\phi}^2 = 2c^2e^{\frac{-2\phi}{\sqrt{k}l_s}}(1+ \Psi e^{\frac{2\phi}{\sqrt{k}l_s}})+2 \sqrt{2} c \big(R+ \tilde{c}(\epsilon_++\epsilon_-)+Y(\epsilon_+-\epsilon_-)\big)-(Y+\tilde{c})^2+2 l_s^2 \left(q_L^2-q_R^2\right).$}
\label{solg2}
\end{eqnarray}
Equations \eqref{solg1} and \eqref{solg2} can be written (using \eqref{ecg}, \eqref{con}, \eqref{Xct}) as
\begin{eqnarray}\label{tdot}
\dot{t}&=&l_s^2E\left[e^{-\frac{2\phi}{\sqrt{k}l_s}}+\Psi\left(1-\frac{E_c}{E}\right)\right],\\
\dot{\phi}^2&=&l_s^4E^2e^{-\frac{2\phi}{\sqrt{k}l_s}}+ \Psi l_s^4E^2 -2l_s^4\Psi E_cE-2l_s^2q_R^2~.\label{phidot}
\end{eqnarray}
Following \cite{Chakraborty:2020swe}, we will take the attitude that physical states in the quantum theory are obtained by quantization of smooth classical trajectories $(\phi(\tau), t(\tau))$, that interpolate between $\phi\to-\infty$ and $\phi\to\infty$ as $t$ goes from $-\infty$  to $\infty$. Thus, we next analyze the constraints on the energy $E$ for given values of the couplings and charges that give rise to such trajectories. As mentioned above, we will restrict to $E>0$, which is the range appropriate for scattering states.

The requirement that $t(\tau)$  \eqref{tdot} is a monotonic function for all $\phi$ implies that one must have
\begin{eqnarray}\label{post}
\Psi\left(1-\frac{E_c}{E}\right)\geq0.
\end{eqnarray}
Reality of $\dot\phi$ \eqref{phidot} for all $\phi$ implies
\begin{eqnarray}
\Psi l_s^4E^2 -2l_s^4\Psi E_cE-2l_s^2q_R^2\ge 0.\label{posphi}
\end{eqnarray}
One can view \eqref{post}, \eqref{posphi} as constraints on the energy $E$ for fixed values of the couplings and charges. To analyze these constraints it is convenient to study separately the cases $\Psi>0$ and $\Psi<0$ (recall the definition of $\Psi$, \eqref{A}).

For $\Psi>0$, the inequality \eqref{post} implies that $E\ge E_c$. For negative $E_c$ \eqref{ecg}, one has instead $E\ge 0$. The condition \eqref{posphi} requires $E$ to lie outside the interval $(E_-,E_+)$ where $E_\pm$,  given by
\begin{eqnarray}\label{epm}
E_\pm=E_c\pm\sqrt{E_c^2+\frac{2q_R^2}{l_s^2\Psi}}~,
\end{eqnarray}
are the roots of the  quadratic polynomial  \eqref{posphi}. $E_-$ is negative and $E_+$ is positive, so taking into account the positivity of $E$ implies that one must have
$E>E_+$. Note that this inequality is stronger than that obtained from \eqref{post} for all $E_c$. Thus, we conclude that the trajectories are smooth and reach the boundary $\phi\to\infty$ for all $E\ge E_+$.

It is instructive to compare this result to the semiclassical spectrum \eqref{mshellg}. There, the energy is parametrized by the radial momentum $P_\phi$. As $P_\phi$ varies from zero to infinity, the energy varies over the range $E_+\le E<\infty$. Thus, we conclude that the semiclassical states \eqref{mshellg} are obtained by quantization of the smooth classical trajectories  \eqref{tdot}, \eqref{phidot}, generalizing the results of \cite{Chakraborty:2020swe}.

For $\Psi<0$, the condition \eqref{post} implies $E\le E_c$. Since $E\ge 0$, this is possible only when $E_c>0$; thus, we will restrict to this case in what follows. The condition \eqref{posphi} is satisfied if the energy lies in the range $E_-\le E\le E_+$. Thus, we conclude that conditions  \eqref{post}, \eqref{posphi} are simultaneously satisfied when $E_-\le E\le E_c$ (note that for $E_c>0, \Psi<0$, $E_-$ is positive).

This result is again compatible with the semiclassical spectrum \eqref{mshellg}. The minimal value of $E$ is $E_-$, attained when $P_\phi=0$. As we increase $P_\phi^2$, $E$ grows. Its maximal value $E=E_c$ corresponds to the $P_\phi$ for which the square root in \eqref{mshellg} vanishes. Thus, the range of energies of the semiclassical states is the same as the range of energies for which the classical trajectory of the string satisfies the conditions listed above.

To see the origin of the upper bound on energies from the classical perspective, it is instructive to examine the trajectories of strings with $E>E_c>0$ for $\Psi<0$. Looking back at \eqref{tdot}, we see that in this case the sign of $\dot t$ changes at a finite value of $\phi$,
\begin{eqnarray}
\phi^\ast  = \frac{\sqrt{k}l_s}{2}\log \left(\frac{E/|\Psi|}{E-E_c}\right).\label{phistarg}
\end{eqnarray}
 At that point $t(\tau)$ turns around and the resulting trajectory becomes multi-valued (a given value of $t$ corresponds to two different $\phi$'s). Furthermore, there is a bound on the value of $t$ for which the string exists. This is very similar to what was found in the $T\bar T$ case in \cite{Chakraborty:2020swe}, and as there we will take the attitude that such trajectories do not give upon quantization (delta function) normalizable states in the quantum theory.

 An interesting feature of the above trajectories is that the location of the point $\phi=\phi^*$ at which $t$ turns around, \eqref{phistarg}, satisfies
\begin{eqnarray}\label{boundphi}
\phi^*\ge -\frac{\sqrt{k}l_s}{2}\log |\Psi|.
\end{eqnarray}
Comparing this to \eqref{cond}, we see that it always lies in the region where closed timelike curves exist. When the energy $E$ exceeds $E_c$ by a small amount, the singularity occurs at very large $\phi$ (\ie\ $\phi^*\to\infty$ as $E\to E_c$).  As the energy increases, $\phi^*$ \eqref{phistarg} decreases, and as $E\to\infty$, the singularity of the trajectory approaches the boundary of the region where CTC's exist (the region complementary to \eqref{cond}).

%\noindent{\bf Black holes}

Another calculation done in \cite{Apolo:2019zai,Chakraborty:2020swe} involves black holes in the deformed background of \cite{Giveon:2017nie}.
These black holes describe the high-energy states in the single-trace $T\bar T$ deformed background corresponding to the current-current deformation of $AdS_3$ mentioned above. Thus, they dominate the thermodynamics in the thermodynamic limit. The thermodynamics of these black holes was found to be
sensible for both signs of the coupling.

For positive coupling, the energy of the black hole is not bounded from above, while the Bekenstein-Hawking temperature is, a reflection of the Hagedorn density of states at asymptotically high energies. For negative coupling, the situation is reversed: the energy is bounded from above, while the temperature is not. As the temperature $T\to\infty$, the corresponding energy approaches the upper bound. The specific heat of the black holes was found to be positive for all accessible temperatures, for both signs of the coupling.

One interesting aspect of the analysis of \cite{Apolo:2019zai,Chakraborty:2020swe} was the qualitative form of the black hole spacetime for negative coupling. As mentioned above, at zero temperature the spacetime has a curvature singularity at a finite value of the radial coordinate. For finite temperature one also has the horizon of the corresponding black hole. The radial position of the horizon increases with the temperature, but the singularity was found in \cite{Apolo:2019zai,Chakraborty:2020swe} to always be outside the black hole. In effect, increasing the size of the black hole pushes the singularity to larger radial position. The horizon approaches the singularity only in the limit when the temperature goes to infinity. This allows one to perform a black hole thermodynamic analysis, whose results are described in \cite{Apolo:2019zai,Chakraborty:2020swe} and mentioned above.

Another interesting result of \cite{Chakraborty:2020swe,Apolo:2019zai} was that the spectrum of black-hole excitations appears to be well described by that of a symmetric product of $T\bar T$ deformed CFT's,
\begin{eqnarray}
\left(\MM_{6k}\right)^p/S_p.\label{symmfull}
 \end{eqnarray}
Here, $\MM_{6k}$ is a CFT with central charge $c=6k$, like $\MM_{6k}^{(L)}$, \eqref{symmm}, but it has more states. More precisely,
the former is a unitary, modular invariant CFT, which contains a normalizable $SL(2,\mathbb{R})$ invariant vacuum,
and thus has a Cardy high-energy density of states with $c=6k$.
The latter is also modular invariant, but it does not contain an $SL(2,\mathbb{R})$ invariant state (like Liouville
theory). Its high-energy density of states is governed by $c_{\rm eff}=6\left(2-\frac{1}{k}\right)$;
see \cite{Giveon:2005mi} for more details.

In the description \eqref{symmfull}, the black holes correspond to generic states at the relevant energy $E$. Since the specific heat of a $T\bar T$ deformed CFT is positive, such states are obtained by dividing the energy roughly equally among the factors of the symmetric product \cite{Giveon:2017nie}. This gives the same thermodynamics as that obtained from the black hole analysis. The fact that this works is intriguing, since in the energy regime described by black holes it is not obvious that the symmetric product structure \eqref{symmfull} should be valid.

It is thus interesting to generalize the black hole analysis of \cite{Chakraborty:2020swe,Apolo:2019zai} to the case of a general linear combination of single-trace
$T\bar T$, $J\bar T$ and $T\bar J$ deformed CFT, in order to see whether the lessons drawn in \cite{Chakraborty:2020swe,Apolo:2019zai} for the $T\bar T$ case are general. In particular, it is interesting to see whether the black hole horizon is located in a regular part of spacetime for all values of the couplings, and whether the corresponding black hole thermodynamics is sensible. It would also be interesting to see whether the description of the black hole states in terms of a symmetric product of the form \eqref{symmfull} generalizes to the full space of deformed theories, and to follow the various qualitative
phenomena, such as the existence of maximal energy and/or temperature, as a function of the couplings.

We expect the picture obtained in \cite{Chakraborty:2020swe,Apolo:2019zai} for the single-trace $T\bar T$ case to generalize to the full space of theories.
Unfortunately, the geometry of generic black holes in the general case is not known;
only a rather small sub-class of them was found.
The black holes in the $T\bar T$ case appear in \cite{Chakraborty:2020swe,Apolo:2019zai}.
In \cite{Chakraborty:2020swe}, they are described in Schwarzschild-like coordinates,
which arise naturally from their description in terms of the corresponding NS5-branes near-horizon,
while in \cite{Apolo:2019zai}, they are presented in isotropic-like coordinates.
A certain class of charged black holes in the $J\bar T$ case --
warped BTZ black holes whose properties are similar to the undeformed BTZ ones, appear in \cite{Apolo:2019yfj}.
It would be interesting to confirm (or disprove) the above expectations in the general case,
but this is beyond the scope of this letter.

%\section{Discussion}

%\appendix

\section*{Acknowledgements}
We thank L. Apolo and W. Song for many helpful discussions.
The work of SC is supported by the Infosys Endowment for the study of the Quantum Structure of Spacetime.
The work of AG and DK is supported in part by BSF grant number 2018068.
The work of AG is also supported in part by a center of excellence supported
by the Israel Science Foundation (grant number 2289/18).
The work of DK is also supported in part by DOE grant de-sc0009924.

\newpage

%\bibliography{ref}\bibliographystyle{JHEP}

\providecommand{\href}[2]{#2}\begingroup\raggedright\endgroup

\end{document}